# Magnetic evolution of Spinel $Mn_{1-x}Zn_xCr_2O_4$ single crystals

G. T. Lin[1,2], X. Luo[1*], Q. L. Pei[1], F. C. Chen[1,2], C. Yang[1,2], J. Y. Song[1,2],

L.H. Yin[1], W. H. Song[1], and Y. P. Sun[3, 1, 4*]

[1] Key Laboratory of Materials Physics, Institute of Solid State Physics, Chinese Academy of Sciences, Hefei, 230031, China

[2] University of Science and Technology of China, Hefei, 230026, China

[3] High Magnetic Field Laboratory, Chinese Academy of Sciences, Hefei, 230031, China

[4] Collaborative Innovation Center of Advanced Microstructures, Nanjing University, Nanjing, 210093, China

## Abstract

$Mn_{1-x}Zn_xCr_2O_4$ ($0 \leq x \leq 1$) single crystals have been grown by the chemical vapor transport (CVT) method. The crystallographic, magnetic, and thermal transport properties of the single crystals were investigated by the room-temperature X-ray diffraction, magnetization $M(T)$ and specific heat $C_P(T)$ measurements. $Mn_{1-x}Zn_xCr_2O_4$ crystals show a cubic structure, the lattice constant $a$ decreases with the increasing content $x$ of the doped $Zn^{2+}$ ions and follows the Vegard law. Based on the magnetization and heat capacity measurements, the magnetic evolution of $Mn_{1-x}Zn_xCr_2O_4$ crystals has been discussed. For $0 \leq x \leq 0.3$, the magnetic ground state is the coexistence of the collinear ferrimagnetic order (CFIM) and spiral ferrimagnetic one (SFIM), which is similar to that of the parent $MnCr_2O_4$. When $x$ changes from 0.3 to 0.8, the SFIM is progressively suppressed and spin glass-like behavior is observed. When $x$ is above 0.8, an antiferromagnetic (AFM) order presents. At the same time, the magnetic specific heat ($C_{mag.}$) was also investigated and the results are coincident with the magnetic measurements. The possible reasons based on the disorder effect and the reduced molecular field effect induced by the substitution of $Mn^{2+}$ ions by nonmagnetic $Zn^{2+}$ ones in $Mn_{1-x}Zn_xCr_2O_4$ crystals have been discussed.

Corresponding author: xluo@issp.ac.cn and ypsun@issp.ac.cn.

# I Introduction

The chalcogenide spinel compounds, named as $ACr_2X_4$ (A=3$d$ transitional metals, Cd and Hg, X=O, S, Se), have attracted special interest in the past ten years because a variety of important physical effects have been found in these compounds, such as colossal magnetocapacitance (MC), multiferroicity, spin frustration and so on.[1-9] Because of the strong coupling among charge, spin and lattice degrees of freedom, $ACr_2X_4$ presents not only interesting phenomena but also complicated magnetic structures.

Among $ACr_2X_4$ compounds, cubic spinel $ACr_2O_4$ oxides have special characters. The unfilled $3d^3$ shells of $Cr^{3+}$ ions form isotropic $S$=3/2 degree of freedom on a lattice of corner-sharing tetrahedron. When the tetrahedral A-site (A = Mg, Zn, Cd and Hg) is occupied by non-magnetic ions, the main magnetic interaction is the strong $J_{CrCr}$ antiferromagnetic (AFM) direct exchange between the nearest-neighbor ions.[5-7] And these compounds show strongly geometrical frustration.[4,8,10-14] On the other hand, when the tetrahedral A-site is occupied by a magnetic ion, such as A=Mn, Co, Fe, and Ni, the magnetic frustration is partially relieved by the $J_{ACr}$ super-exchange interaction.[5,15] The $A^{2+}$-O-$Cr^{3+}$ interaction is usually collectively stronger than the frustrated ones between the $Cr^{3+}$ ions.[16] In this case, the system presents nearly degenerated ground states and it develops complex low temperature magnetic order. Among the spinel $ACr_2O_4$ compounds, $MnCr_2O_4$ and $ZnCr_2O_4$ are two typical ones. In $MnCr_2O_4$, the collinearly ferrimagnetic (CFIM) temperature $T_C$ is observed around 41-51 K, which are dependent on the polycrystalline samples and single crystals. In addition, the sample exists a characteristic temperature $T_S$. As $T < T_S$, the long-range FIM and the short-range spiral FIM (SFIM) coexists. Between $T_C$ and $T_S$, the long-range FIM with an easy axis parallel to the <1 1 0> direction occurs.[7,9,17-19] Very recently, the multiferroicity has also been reported in $MnCr_2O_4$ below the $T_S$.[9] However, $ZnCr_2O_4$ shows strikingly different characters, such as strongly geometrical frustration (the frustration factor $f \approx 31$) and high Curie-Weiss temperature $\theta$ = 390-400 K. The AFM with the spin-Jahn-Teller distortion, which favors a relief of the geometrical frustration, appears around $T_N$=12 K with the character of a first-order phase transition.[8,20,21] From above reported works, it seems to mean that the molecular field of the A sites can be effectively tuned and has the important effect on the ground state of the spinel oxides. Because the emergent phenomena present in spinel $MnCr_2O_4$ and $ZnCr_2O_4$ compounds, the magnetic evaluation of $Mn_{1-x}Zn_xCr_2O_4$ oxides are really deserved to be investigated. Although few work has been done on the $Zn^{2+}$ ions doped $MnCr_2O_4$ compounds, the comprehensive study is still missing and the evolution of magnetic ground state is not very clear.[22,23] In order to further understand magnetic evolution of the ground state, herein, we investigate the effect of non-magnetic $Zn^{2+}$ ions doping at the magnetic A sites of $Mn_{1-x}Zn_xCr_2O_4$ single crystals. The magnetic phase diagram of $Mn_{1-x}Zn_xCr_2O_4$ single crystals is obtained. We also discussed the magnetic evolution based on the disorder effect and the reduced molecular field one induced by the substitution of $Mn^{2+}$ ions by nonmagnetic $Zn^{2+}$ ones in $Mn_{1-x}Zn_xCr_2O_4$ crystals.

## II Experimental results

$Mn_{1-x}Zn_xCr_2O_4$ single crystals were grown by the chemical vapor transport (CVT) method, with $CrCl_3$ powders as the transport agent. Firstly, polycrystalline $Mn_{1-x}Zn_xCr_2O_4$ were made by the solid state reaction. The stoichiometric amounts of $Cr_2O_3$ (99%, Alfa Aesar), ZnO (99.9%, Alfa Aesar) and MnO (99%, Alfa Aesar) powders were mixed in air and sintered at 1300℃ for 20 h for several times. Powder X-ray diffraction (XRD) at room temperature revealed a single-phase without any detected impurities. Secondly, polycrystalline $Mn_{1-x}Zn_xCr_2O_4$ mixed with the $CrCl_3$ powders were ground again and sealed into several quartz tubes. All were done in the Ar-filled glove-box. All sealed quartz tubes were put in a two-zone tube furnace. The hot side is about 1060℃ and the cold side is about 1015℃, and dwelled for 10 days, then slowly cooled down to room temperature with a rate of 15℃/h. The crystals are octahedral with shining surfaces and the size is about 1.2*1.2*1.2 mm$^3$. Heat capacity was measured using the Quantum Design physical properties measurement system (PPMS-9T) and magnetic properties were performed by the magnetic property measurement system (MPMS-XL5).

## III Results and discussion

Figure 1 (a) shows the XRD pattern of $Mn_{0.8}Zn_{0.2}Cr_2O_4$ powder obtained by crushing the single crystals, which is selected as a typical sample. It also presents the structural Rietveld refinement profiles of the XRD data by the Highscore software. The refinements of the XRD data indicate that the crystals are single-phase since no extra peaks were observed. It is found that the crystals belong to the normal spinel structure and the space group is $Fd\bar{3}m$ (space group No. is 227). X-ray diffraction patterns of $Mn_{1-x}Zn_xCr_2O_4$ powder are presented in Fig. 1(c). It also presents a single-phase for all samples. The crystal structure of these samples does not change with the Zn doping content $x$ at room temperature, the lattice parameter $a$ decreases from 0.844 nm for $x$=0 to 0.833nm for $x$=1.0, which confirms that the partial $Mn^{2+}$ sites are occupied by the $Zn^{2+}$ ions. The reduced lattice parameter is related to the fact that the $Zn^{2+}$ ion radius (0.06 nm) is smaller than that of $Mn^{2+}$ one (0.066 nm). The lattice constant $a$ for all samples follows the Vegard law. On the other hand, the positions of Bragg peak (511) (shown in Fig. 1 (d)) move to high degree with the increasing content $x$, which well agrees with the Rietveld refinement results.

In order to investigate the macroscopic magnetic properties of $Mn_{1-x}Zn_xCr_2O_4$ single crystals, we carried out the measurement of the magnetization $M(T)$ as the function of temperature. Figure 2(a)-(f) show the $M(T)$ under the zero field-cooled (ZFC), field-cooled (FCC) and field-warming (FCW) modes with the applied magnetic field parallel to the <111> direction for $Mn_{1-x}Zn_xCr_2O_4$ single crystals, respectively. Fig. 2 (a) and (b) show a FIM behavior for $x$≤0.2. A collapse of the cusp is observed for the compounds with higher $x$ in Fig. 2 (c), which indicates the magnetic ground state is changed by the $Zn^{2+}$ ion doping, namely, the SFIM may be suppressed and the magnetic ground state changes into the spin glass state in the spinel oxides.[9,24] For x≥0.6, compared with the lower doped content $x$, the value of the magnetization $M$ is much smaller and the ZFC and FCC curves are obviously irreversibility in $x$=0.6 and 0.8 (as shown in Fig. 2 (d) and (e)). It may mean that both AFM and FIM orders are perturbed, then destroyed and eventually the spin glass of the ground state presents.[24] In Fig. 2 (f), accompanied by the sharp drop of magnetization $M(T)$, the AFM order occurs at $T_N$=12 K, which is in agreement with the reported data. It is related to a structural phase transition from cubic $Fd\overline{3}m$ phase to the tetragonal $I4_1/amd$ one at $T_N$ for $ZnCr_2O_4$.[6,25] In order to further investigate the nature of the magnetic structure of $Mn_{1-x}Zn_xCr_2O_4$ single crystals, the magnetic field dependence of the magnetization ($M(H)$) for all crystals at $T$=5 K are shown in Fig. 3. Except for the parent $MnCr_2O_4$, it shows that the coercivity $H_C$ increases with the increasing $x$ for $x$≤0.6. The saturated magnetization $M_S$ is nearly decreasing with increasing content $x$.

Now we focus on the nature of Zn doped $MnCr_2O_4$ single crystals, we did the analysis on the temperature dependent inverse susceptibility $\chi^{-1}(T)$. Firstly, we pay attention to the parent compound $MnCr_2O_4$, the FIM order $T_C$ and SFIM one $T_S$ are 52.7 K and 24.4 K obtained from the peak of the heat capacity $C_P/T$, respectively. From the mean-field theory, for a AFM system and FIM one, the temperature dependent inverse susceptibility above $T_C$ can be described by the Curie-Weiss law (Eq.(1)) and the hyperbolic behavior characteristic of ferrimagnets (Eq.(2)):[7,26]

$$\frac{1}{\chi} = \frac{T-\theta}{C} \qquad (1)$$

$$\frac{1}{\chi_{FIM}} = \frac{T-\theta}{C} - \frac{\zeta}{T-\theta'} \qquad (2)$$

where $C$ is the Curie constant, $\theta$ is the Weiss temperature. In Eq. (2), the first term is the hyperbole high-T asymptote that has a CW form and the second term is the hyperbole low-T asymptote. All the fitting results are summarized in Fig. 4 (b). The substitution of $Mn^{2+}$ ions by $Zn^{2+}$ ions progressively perturbs the FIM structure, leading to a reduction of the effective magnetic moment $\mu_{eff}$ ($\mu_{eff} = \sqrt{8C}\mu_B$) and magnetic fraction factor $f$ ($f=\theta/T_N$) for increasing $x$. The tetragonal position occupied by The non-magnetic $Zn^{2+}$ ions is responsible for the above results, which are consistent with the ones obtained from the above $M(T)$ and $M(H)$ measurements.

To study the thermal property, we performed the detailed analysis on the temperature dependent specific heat $C_P/T$ of $Mn_{1-x}Zn_xCr_2O_4$ single crystals, as shown in Fig. 5 (a). For $ZnCr_2O_4$, which yields a sharp specific heat anomaly with the character of the first-phase transition at $T_N$=12.2 K. With the decrease of $x$, this heat capacity anomaly is clearly suppressed. For 0.6≤$x$≤0.8, the heat capacity peak disappears and the specific heat presents a monotonous smooth curve, which shows a spin glass behavior.[25] It agrees well with the magnetic results. When $x$≤0.4, just one specific heat peak is observed, as present in the inset of Fig. 5 (a), which is related to the CFIM. When $x$ continues to decrease, two specific heat peaks are observed for 0≤$x$≤0.2, and which are corresponding to the CFIM and SFIM orders. As $x$≤0.2, it is obvious that the magnetic structure of $Mn_{1-x}Zn_xCr_2O_4$ samples in the FIM regions is in good agreement with $MnCr_2O_4$. In addition, as shown in Fig. 5 (b), the low-temperature magnetic specific heat presents linear variation in $Mn_{1-x}Zn_xCr_2O_4$. The linear variation suggests a constant density of states of the low-temperature magnetic excitations, which is claimed to be a common feature of spin glasses.[27,28] For 0.4≤$x$≤0.8, a broad magnetic specific heat anomaly is observed for spin glasses, indicating that short-range-order contributions extend up to very high temperatures. Figure 5 (b) shows the temperature dependence of the magnetic heat capacity $C_{mag.}$ for $Mn_{1-x}Zn_xCr_2O_4$. Since all the samples show insulating behavior, we can ignore the electronic contribution to the heat capacity, the $C_{mag.}$ can be calculated by the following equations:[26]

$$C_{VDebye}(T) = 9R\left(\frac{T}{\Theta_D}\right)^3 \int_0^{\Theta_D/T} \frac{x^4 e^x}{(e^x - 1)^2} dx \tag{3}$$

$$C_{mag.}(T) = C_p(T) - nC_{VDebye}(T) \tag{4}$$

where $n = 7$ is the number of atoms per formula unit, $R$ is the molar gas constant and $\Theta_D$ is the Debye temperature. The sum of Debye functions accounts for the lattice contribution to specific heat. We can get the $C_{mag.}$ by Eq. (4). We also can obtain the magnetic entropy $S_{mag.}$ from the $C_{mag.}$, which is calculated by integral of the $C_{mag.}/T$ versus $T$: [26]

$$S_{mag.}(T) = \int_0^T \frac{C_{mag.}(T)}{T} dT \tag{5}$$

The $T$ dependence of $S_{mag.}$ is shown in Fig. 6. $S_{mag.}$ is monotonously decreasing with the increasing content $x$ except for $ZnCr_2O_4$. In addition, the change of Debye temperature behaves as firstly decreasing and then increasing with the increasing content $x$ except for $ZnCr_2O_4$, the $x=0.4$ sample has a minimum value $\Theta_D=270K$. The abnormality of $ZnCr_2O_4$ may be attributed to the character of the first-phase transition at $T_N=12.2$ K.

Based on our obtained results, we summarize the magnetic phase diagram of $Mn_{1-x}Zn_xCr_2O_4$ single crystals as plotted in Fig. 7. Now, let us try to understand the magnetic evolution in $Mn_{1-x}Zn_xCr_2O_4$ single crystals. As we know, the substitution of $Mn^{2+}$ ($S=5/2$) by $Zn^{2+}$ ($S=0$) ions usually can induce following effects: the shrinkage of lattice, the disorder of A sites and the decreased content of magnetic $Mn^{2+}$ ions. As shown in Fig. 2, for $x \leq 0.4$, it can be seen that all the samples undergo a transition from PM to FIM. The ZFC and FCC curves are obviously irreversibility in 100Oe, which can be attributed to magnetic frustration or a transition into a spin-glass phase.[29-31] We note that the magnetization in 100Oe decreases and the transition temperature $T_C$ is lower than that of $MnCr_2O_4$ with the increase of $x$. Meanwhile, the FIM order presents an easy axis along the $[1\bar{1}0]$ direction in $MnCr_2O_4$ below $T_C$, which is identical with the magnetic moment direction of $Mn^{2+}$ ion.[7,32] Although the shrinkage of lattice could enhance the exchange interaction between Cr–Mn ions via oxygen *2p* orbits, the decreased content of Mn ions is the major factor, which is responsible for the above observed phenomenon.[33] In MnCr$_2$O$_4$, the magnetization of the tetrahedral A (Mn$^{2+}$) site is antiparallel to that of the octahedral B (Cr$^{3+}$) site. In the light of molecular field model, the A-O-B super exchange interactions predominate the A-A and B-B interactions. However, as it is shown in Fig. 3, the abnormality of MnCr$_2$O$_4$ may be attributed to the cation distribution of A and B sites. [29] This is because Mn$^{2+}$ site is slightly occupied by Cr$^{3+}$ ions, which leads to the enhanced coercivity and the reduced magnetization.[34] With the substitution of magnetic ions in A site by the Zn$^{2+}$ (has preferentially A-site occupancy), the cation distribution of A and B sites will be consistent with that of normal spinel structure. For $x \geq 0.05$, the replacement of Mn$^{2+}$ with non-magnetic Zn$^{2+}$ leads to the weakening of the A-O-B super exchange interaction. This would further disturb the magnetic couplings and lead to a reduction of the magnetization. The substitution of Zn$^{2+}$ ion for Mn$^{2+}$ or Co$^{2+}$ ions has relatively similar physical properties. For example, Brent C Melot et al.[24] has reported magnetic phase evolution in the spinel compounds Co$_{1-x}$Zn$_x$Cr$_2$O$_4$, which is similar to our results obtained in Mn$_{1-x}$Zn$_x$Cr$_2$O$_4$. At the same time, the structure of Co$_{1-x}$Zn$_x$Cr$_2$O$_4$ at low temperature (*T*=5 K) is still modeled by the cubic space group $Fd\bar{3}m$ for x$\leq$0.9.[25] For $x \geq 0.6$, the magnetic coupling interactions become more complicated. The Mn-O-Cr superexchange interaction partially breaks the spin degeneracy of the ground state and the coupling between the spin and lattice degrees of freedom becomes weaker than that of ZnCr$_2$O$_4$ in Mn$_{1-x}$Zn$_x$Cr$_2$O$_4$. It is reasonable that the Mn-O-Cr superexchange interaction could disrupt the coherency of Cr-Cr exchange coupling paths, and then inhibit the spin-Jahn-Teller distortion in Mn$_{0.2}$Zn$_{0.8}$Cr$_2$O$_4$ and Mn$_{0.6}$Zn$_{0.4}$Cr$_2$O$_4$.[25] However, more detail structural experiments at low temperature, including magnetic structure determined using neutron scattering method, are needed in future.

## IV Conclusion

From the room-temperature X-ray diffraction, magnetization *M(T)* and specific heat *C$_P$(T)* measurements, the crystallographic, magnetic, and thermal transport properties of the Mn$_{1-x}$Zn$_x$Cr$_2$O$_4$ single crystals were obtained. The lattice constant *a* decreases with the increasing content *x* of the doped Zn$^{2+}$ ions and follows the Vegard law. The magnetic evolution of Mn$_{1-x}$Zn$_x$Cr$_2$O$_4$ crystals based on the magnetization and specific heat measurements has been discussed. For $0 \leq x \leq 0.3$, the magnetic ground state is the coexistence of CFIM and SFIM, which is similar to that of the parent MnCr$_2$O$_4$. $x$ changes from 0.3 to 0.8, the SFIM is progressively suppressed and spin glass-like behavior is observed. While $x$ is above 0.8, an AFM order presents. The magnetic specific heat ($C_{mag.}$) was also calculated and the results are coincident with the magnetic measurements. The possible reasons based on the disorder effect and the reduced molecular field effect induced by the substitution of Mn$^{2+}$ ions by nonmagnetic Zn$^{2+}$ ones in Mn$_{1-x}$Zn$_x$Cr$_2$O$_4$ crystals have been discussed.

# Acknowledgements

This work was supported by the Joint Funds of the National Natural Science Foundation of China and the Chinese Academy of Sciences' Large-Scale Scientific Facility under contracts U1432139, U1532152, the National Nature Science Foundation of China under contracts 51171177, 11404339, and the Nature Science Foundation of Anhui Province under contract 1508085ME103.

**Figure 1:**

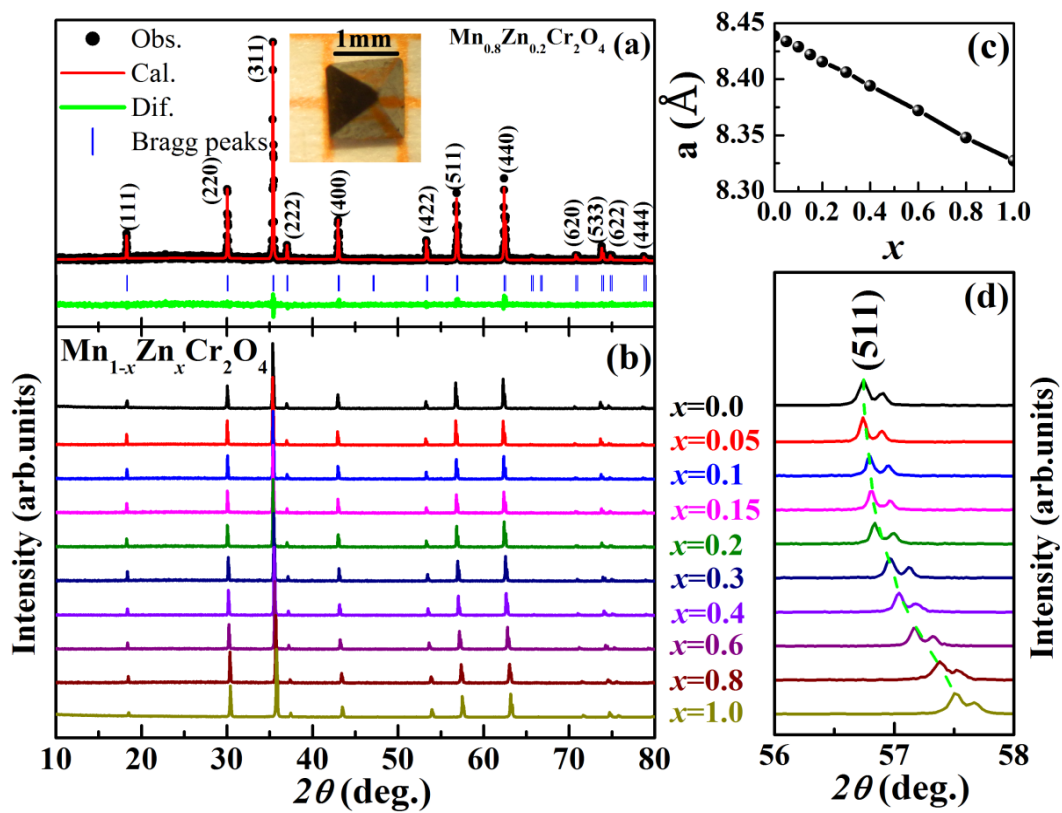

**Fig. 1 (color online): (a)** The refined powder XRD patterns at room temperature for the $Mn_{0.8}Zn_{0.2}Cr_2O_4$ powders crushed by single crystals. The black dots are the experimental data and the red line is the fitting result. The solid line (green line) at the bottom corresponds to the difference between experimental and calculated intensities. The blue bars are the Bragg positions. The inset shows the typical crystal of $Mn_{0.8}Zn_{0.2}Cr_2O_4$, the scale bar is 1 mm; **(b)** Powder XRD patterns for $Mn_{1-x}Zn_xCr_2O_4$ powders crushed by the single crystals; **(c)** The content $x$ of doped $Zn^{2+}$ ions dependence of the fitted lattice parameter $a$; **(d)** The doping amount $x$ of $Zn^{2+}$ ions dependence of the position of Bragg peak (511).

**Figure 2:**

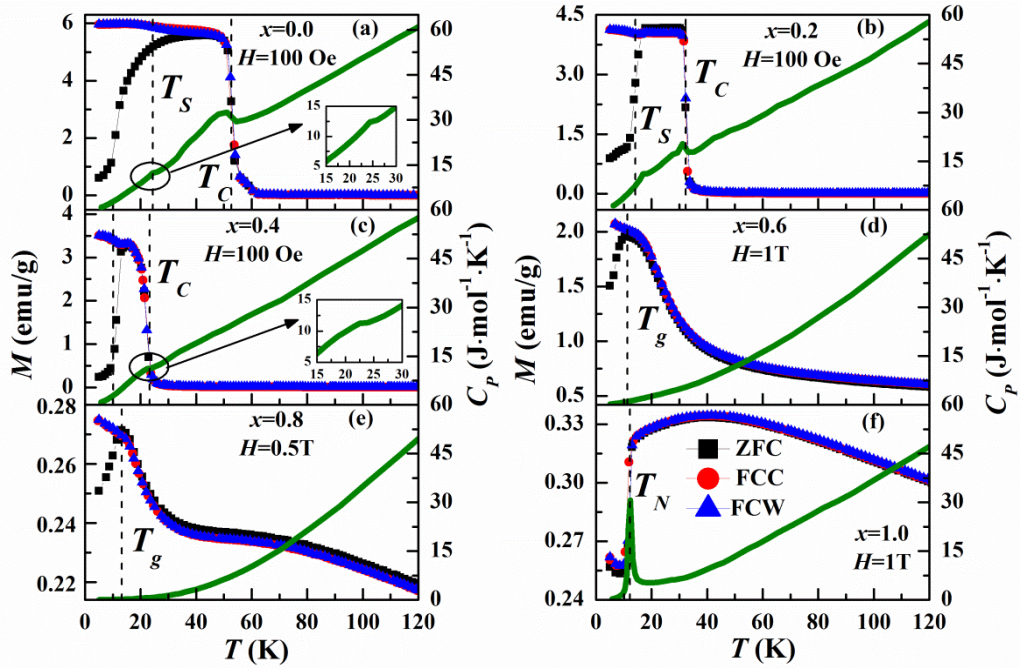

**Fig. 2 (color online): (a)-(f)**: The temperature dependence of magnetization *M(T)* and specific heat $C_P(T)$ for $Mn_{1-x}Zn_xCr_2O_4$ (0≤ x≤1) single crystals. Black, red and blue lines are the magnetization measured under the ZFC, FCC and FCW modes, respectively. $T_C$ are defined as the temperature of the minimum slope of the *M(T)* data under the ZFC modes. Except for $MnCr_2O_4$, $T_S$ are defined as the temperature of the maximum slope of the *M(T)* data under the ZFC modes. $T_S$ of $MnCr_2O_4$ is obtained from the peak of the heat capacity $C_p/T$. The glass freezing temperature $T_g$ is defined as the temperature of maximum magnetization in *M-T* curves. The olive line represents the specific heat data at *H*=0 Oe. Insets of **(a)** and **(c)** present the enlarged view of the specific heat data around $T_C$ or $T_S$.

**Figure 3:**

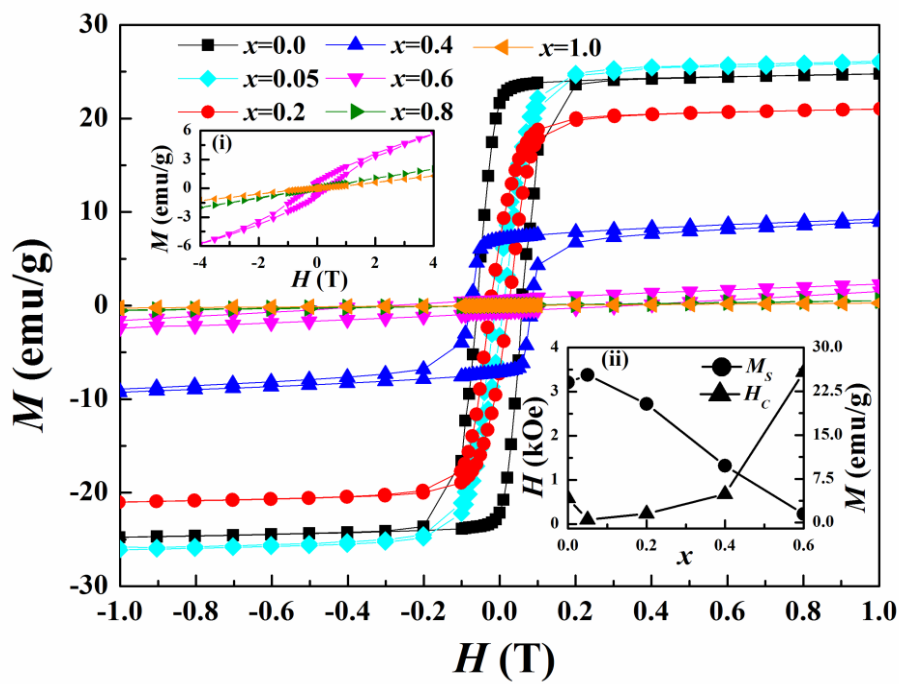

**Fig. 3 (color online):** The magnetic field dependence of magnetization *M(H)* at *T*=5 K for $Mn_{1-x}Zn_xCr_2O_4$ single crystals. The applied magnetic field is along the [111] direction. The left inset presents the enlarged view of the magnetic hysteresis loop for the doping level *x*>0.6. The right inset shows the content *x* of $Zn^{2+}$ ions dependence of the coercivity $H_C$ and the saturation magnetization $M_S$ for $Mn_{1-x}Zn_xCr_2O_4$ single crystals at *T*=5 K.

**Figure 4:**

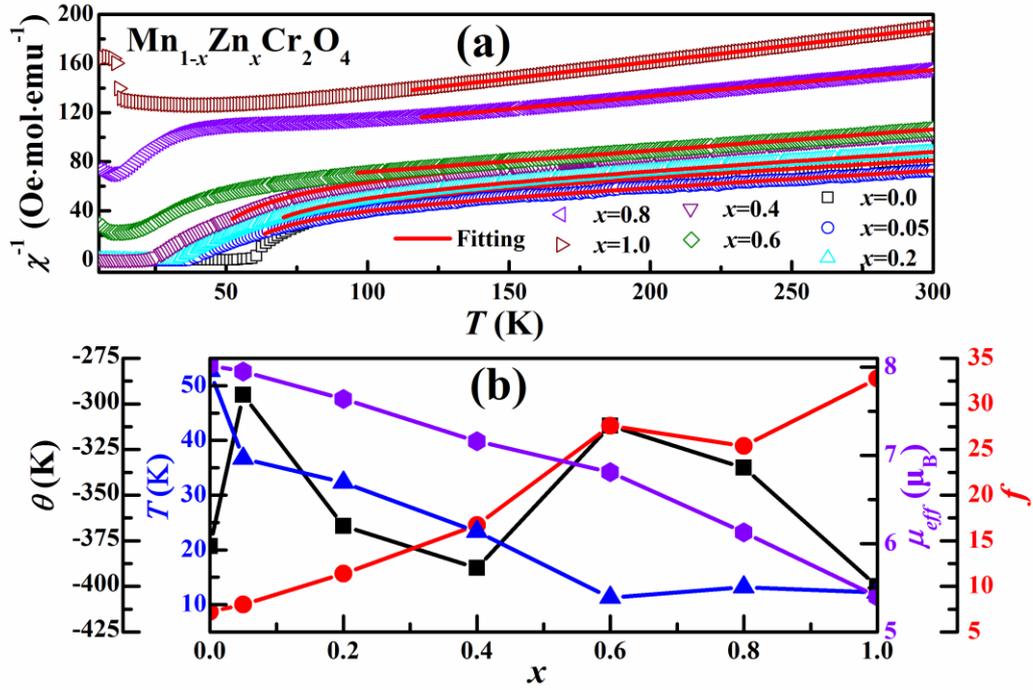

**Fig. 4 (color online): (a)** The inverse susceptibility dependence of the temperature for $Mn_{1-x}Zn_xCr_2O_4$ single crystals. The solid lines are the fitting results according to Eq.(1) and Eq.(2); **(b)** The obtained effective moment $\mu_{eff}$, Weiss temperature $\theta$, the magnetic order temperature $T$ (including $T_C$ and $T_N$) and the frustration factor $f$ dependence of the doping level $x$ of the $Zn^{2+}$ ions.

**Figure 5:**

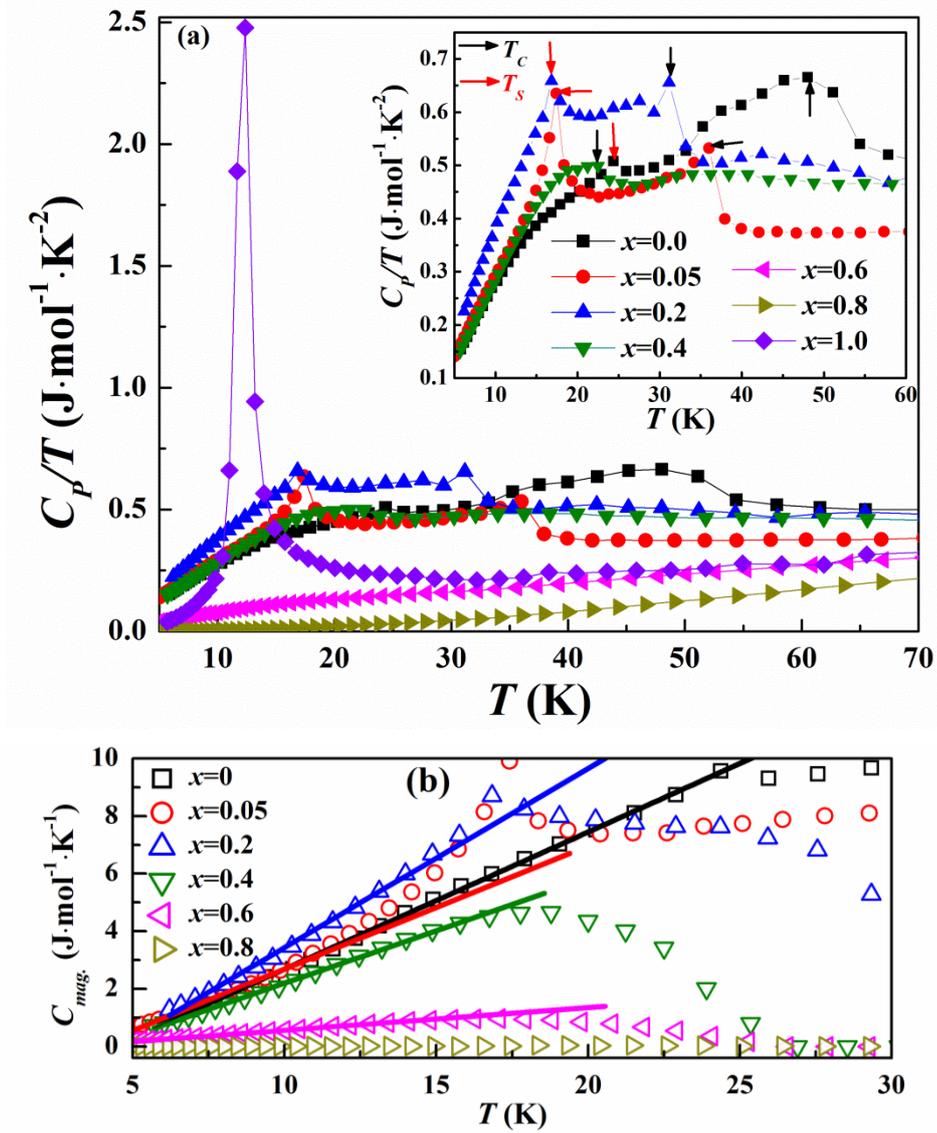

**Fig. 5 (color online): (a)** $C_P/T$ curves of $Mn_{1-x}Zn_xCr_2O_4$ single crystals. The inset presents partial enlarged detail for $0 \leq x \leq 0.4$. The red and blue arrows are corresponding to $T_C$ and $T_S$, respectively; **(b)** shows the linear dependence of magnetic specific heat $C_{mag.}$ at low temperature. The straight line guides the linear dependence.

**Figure 6:**

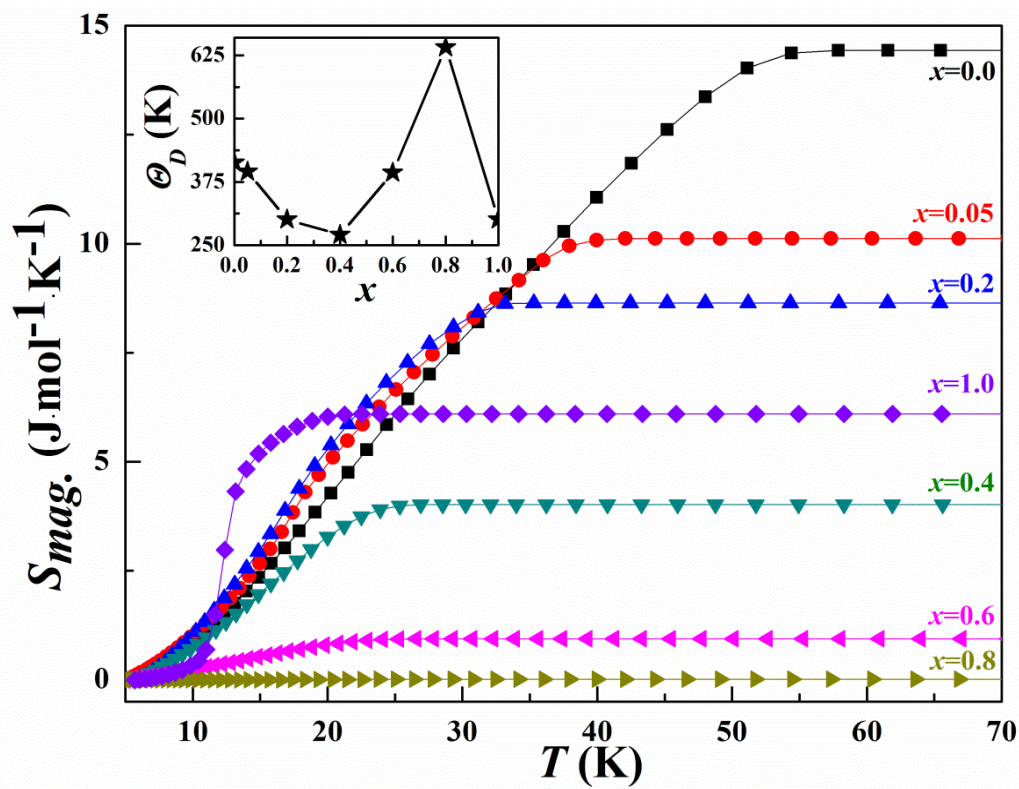

**Fig. 6 (color online):** The temperature dependence of the magnetic entropy $S_{mag.}$ of $Mn_{1-x}Zn_xCr_2O_4$ single crystals. The inset shows the doping content $x$ dependence of the Debye temperature $\Theta_D$.

**Figure 7:**

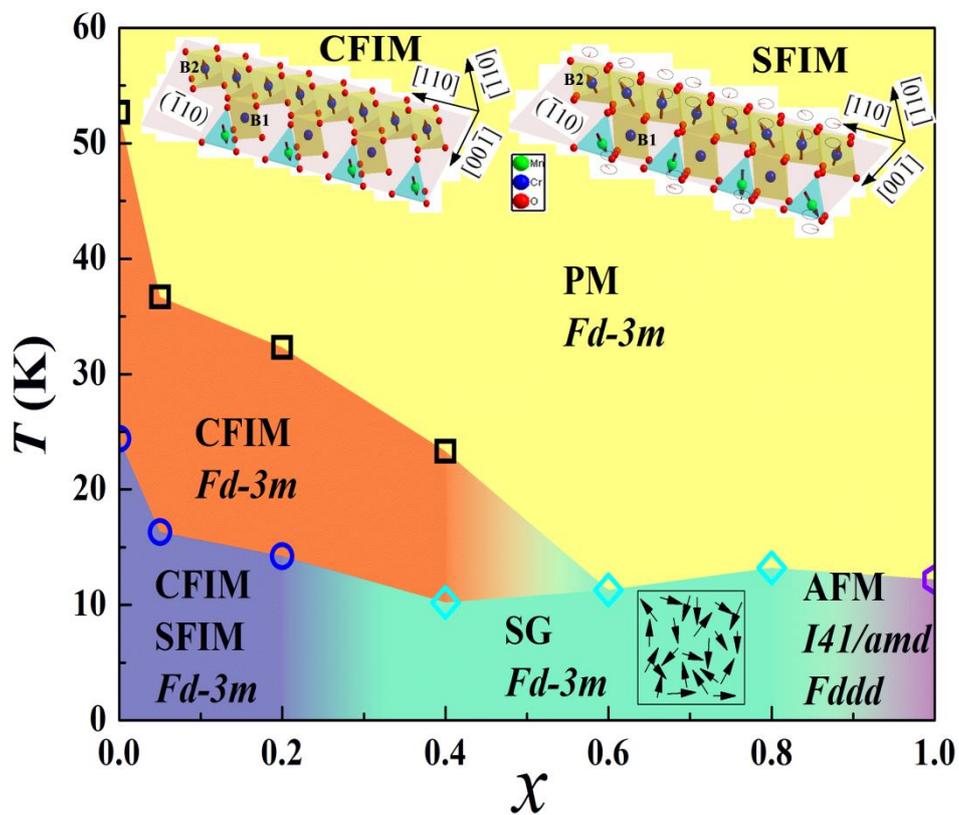

**Fig. 7 (color online):** The magnetic evolution of $Mn_{1-x}Zn_xCr_2O_4$ single crystals. PM, SFIM, CFIM, SG and AFM are corresponding to the paramagnetism, spiral ferrimagnetism, collinear ferrimagnetism, spin glass and antiferromagnetism, respectively.